# Dynamic motion of polar skyrmions in oxide heterostructures

Lizhe Hu[1], Yongjun Wu[1,2*], Yuhui Huang[1], He Tian[3], Zijian Hong[1, 2, *]

[1] School of Materials Science and Engineering, Zhejiang University, Hangzhou 310027, China

[2] State Key Laboratory of Silicon and Advanced Semiconductor Materials, Zhejiang University, Hangzhou, Zhejiang 310027, China

[3] Center of Electron Microscopy, School of Materials Science and Engineering, Zhejiang University, Hangzhou, 310027, China

Email: yongjunwu@zju.edu.cn (Y.W.); hongzijian100@zju.edu.cn (Z.H.)

**Abstract**

Polar skyrmions have been widely investigated in oxide heterostructure recently, due to their exotic properties and intriguing physical insights. Meanwhile, so far, the external field-driven motion of the polar skyrmion, akin to the magnetic counterpart, has yet to be discovered. Here, using phase-field simulations, we demonstrate the dynamic motion of the polar skyrmions with integrated external thermal, electrical, and mechanical stimuli. The external heating reduces the spontaneous polarization hence the skyrmion motion barrier, while the skyrmions shrink under the electric field, which could weaken the lattice pinning and interactions between the skyrmions. The mechanical force transforms the skyrmions into *c*-domain in the vicinity of the indenter center under the electric field, providing the space and driving force needed for the skyrmions to move. This study confirmed that the skyrmions are quasi-particles that can move collectively, while also providing concrete guidance for the further design of polar skyrmion-based electronic devices.

**Introduction**

The recent discovery of exotic ferroelectric topological structures [1-3] such as polar vortex [4-6] and polar skyrmion [7-9], etc., has attracted tremendous attention recently in the condensed matter physics and materials science community. These structures usually have spatially inhomogeneous polarization distributions and exhibit topological protection under externally applied stimuli before the full destruction of the structures. One particular example is the polar skyrmion with whirl-like polarization distributions and a topological charge of ±1 (defined as the surface integration of the Pontryagin density), which has been observed and studied extensively in the $PbTiO_3/SrTiO_3$ (PTO/STO) superlattice on an STO substrate, akin to the ferromagnetic counterpart [8-10]. Many exotic properties have been discovered in polar skyrmions, including negative capacitance [10, 11], chirality [9], unusual topological phase transitions [12], etc.

In a ferromagnetic system, it is now widely accepted that the electric current can drive the directional motion of the magnetic skyrmions [13-22], owing to the spin-transfer torque [14-19] or the spin Hall effects [20, 21]. For instance, Jiang *et al.* have demonstrated current-driven transformation from stripe domains to magnetic skyrmion bubbles in the CoFeB system, which can further move with the electric current after creation [22]. A step further, device applications have been proposed based on the controlled dynamic motion of magnetic skyrmions [23], including transistors [24], logic gates [25, 26], and neuromorphic computing devices [27, 28]. Meanwhile, although the electric field driven creation, annihilation, and expansion have been demonstrated for the polar skyrmions [8, 10, 12, 29], the dynamic motion of the polar skyrmion has yet to be discovered. In general, there are two main obstacles to the motion of skyrmions in a ferroelectric system: (1) the lacking of a mechanism similar to the spin-transfer torque or spin Hall effects; (2) the strong lattice pinning and mechanical constraint in an oxide heterostructure.

Herein, using phase-field simulations, we demonstrated the dynamic, collective motion of the polar skyrmions in a PTO/STO superlattice system, with combined external mechanical, electrical, and thermal stimuli. The application of external heating could reduce the spontaneous polarization and skyrmion moving barrier; the applied electric field could weaken the lattice pinning and interactions between the skyrmions; while the mechanical forces generated by a moving indenter could not only create a "blank" region with enough space and driving force for the motion of the polar skyrmions. It is observed that the skyrmions in the vicinity of the blade-like indenter will move with the motion of the indenter at elevated temperatures and externally applied electric field, with a moving speed of ~5 m/s. We hope this study to spur further interest in the manipulation and motion of the polar skyrmions for the design of polar skyrmion-based electronic devices.

**Main**

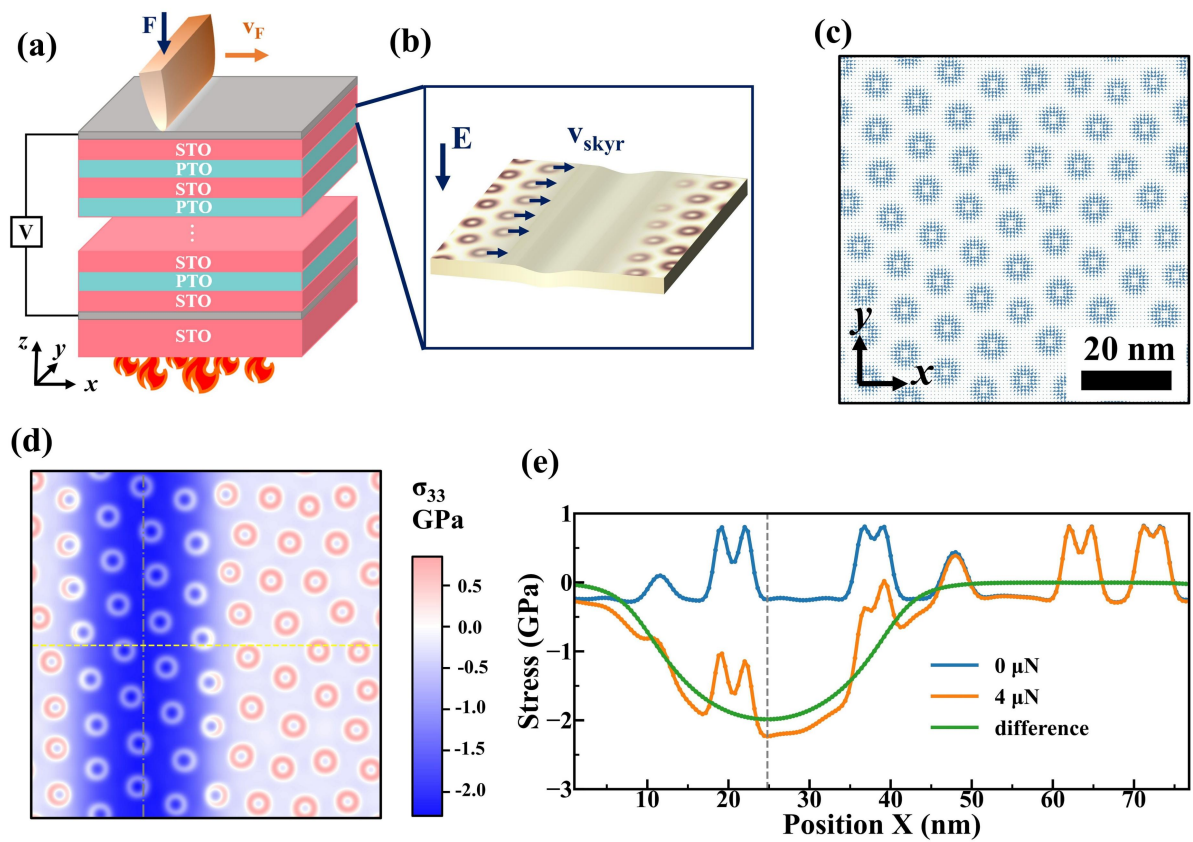

**Fig. 1| Initial setup of the simulation system.** (a) Schematics of the PTO/STO superlattice system under an extra mechanical load and a planar electric potential at elevated temperatures. (b) A banded single $c$-domain is formed underneath the indenter to allow the

skyrmion motion. (c) The planar view of the in-plane polarization at the top of the PTO layer under an electric potential of 9 V, at 500K, shows the formation of stable skyrmions without external mechanical forces. (d) Distribution of stress component $\sigma_{33}$ in the planar view at the top of the PTO layer under 4 μN load under 9 V at 500 K. The grey dash-dot line represents the center of the blade. (e) The comparison of stress component $\sigma_{33}$ with and without external mechanical force along the yellow line in (d).

Details of the phase-field simulations are given in Methods and previous reports [6, 8, 12]. The initial setup for the simulation system is given in **Fig. 1**. The $(PTO)_{16}/(STO)_{16}$ superlattice epitaxially grown on top of an STO substrate (Fig. 1a) is employed as the model system. Previously, the polar skyrmion has been observed in this system [9, 10]. A homogeneous out-of-plane electric field is then applied through the top and bottom electrodes. To reduce the polarization magnitude and skyrmion correlations, the system is heated at an elevated temperature. The mechanical force is further applied through a blade-like indenter, which moves along the in-plane $x$ direction. The schematics of the applied field and skyrmion bubble moving direction are highlighted in Fig. 1(b). The in-plane polarization after applying an electric bias of 9 V at 500 K and zero applied force shows that the center-divergent polar skyrmion is stable, with a shrinking of the bubble size and an increase of the bubble distance as compared to the original polar skyrmion structure at 300 K and 0 V (Fig. 1c). Fig. S1(a) shows the model of the indenter, which is approximated by a semi-cylinder with a radius $R$ at the bottom in contact with the oxide surface. The 2-D surface contact between the tip and film is illustrated in Fig. S1(b), where the blue-filled rectangle is the contact area, the dash-dot line is the central axis of the tip, $L$ is the length of the tip, and $a$ is the half-width of the contact area. The stress distribution under the tip can be approximated by a cylindrical punch indenter model [30, 31]: $\sigma_{33} = -\frac{2F}{\pi a L}\sqrt{1-\frac{r^2}{a^2}}$ $(r \leq a)$, where $F$ is the applied load, and $r$ is the distance to the center of the indenter. The distribution of stress component $\sigma_{33}$ after applying a 4 μN mechanical force at 500 K is plotted in Fig. 1(d), showing a stripe-like high compressive stress region in the vicinity of the indenter center. The line plots of the $\sigma_{33}$ with and without external load along the yellow dot line are depicted in Fig. 1(e), demonstrating that the maximum externally applied out-of-plane stress reaches ~2 GPa under an external force of 4 μN, and decays exponentially away from the center of the indenter. For comparison, the planar view of the skyrmions at the top PTO layer without mechanical force, the stress component $\sigma_{33}$ after applying a 4 μN force, and the line plots of the $\sigma_{33}$ with and without external load at 300 K without external electric potential are plotted in Fig. S1(c)-S1(e), showing a similar trend to the case at 500 K, while the skyrmions are more compact.

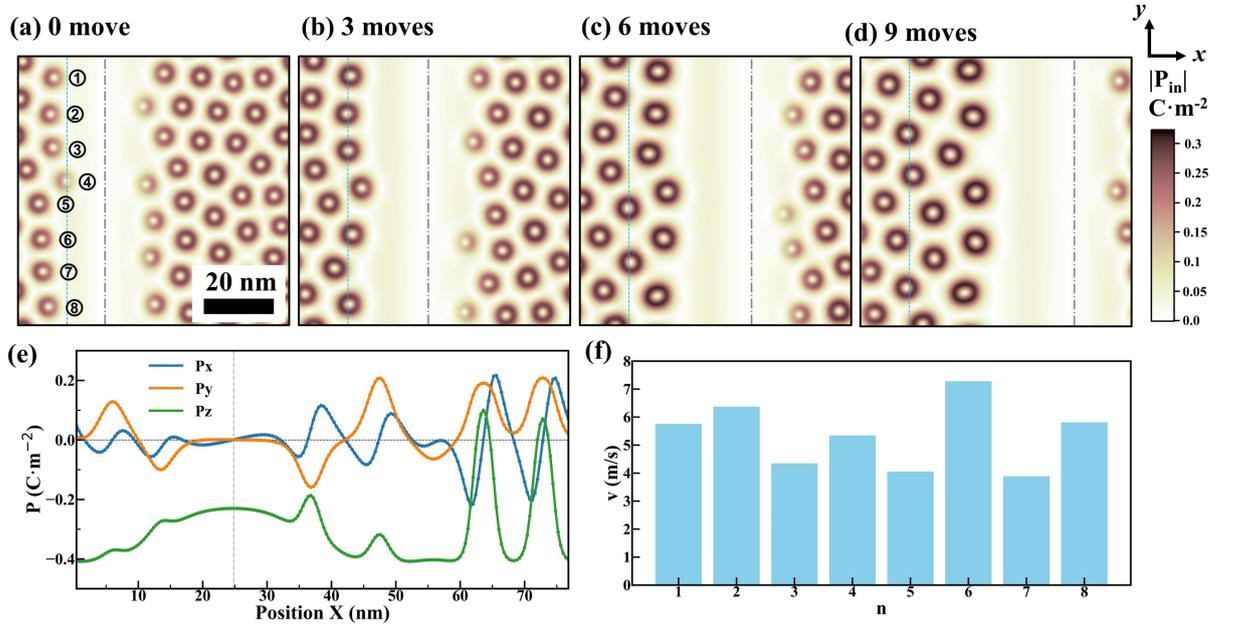

**Fig. 2| The dynamic motion process of polar skyrmions.** (a-d) The dynamic process of polar skyrmions with a moving indenter after 0, 3, 6, and 9 moves, respectively. The grey dash-dot lines represent the central axis of the mechanical indenter at different steps. (e) Line plots of the different polarization components after 0 moves. (f) The average velocity of different skyrmions along the *x* direction.

We first investigate the dynamic motion of polar skyrmions with a moving indenter at 500 K, as shown in **Fig. 2**. In this example, we apply a constant bias of 9 V on the top PTO layer, and a 4 μN mechanical force is simultaneously applied and moved rightward to drive the motion of the skyrmions. Initially, the center of the indenter is located in the gray dashed line, the polar structure under the external mechanical force in equilibrium is plotted in Fig. 2(a). It can be observed that when a specific force is applied on the surface, the skyrmions in the vicinity of the tip are fully switched, forming a single *c*-domain region that can enable the skyrmion motion without strong skyrmion-skyrmion pinning effects. Then, the blade-like indenter moves 4 nm along *x* direction every 20000 timesteps for a total of 36 nm after 9 moves. The dynamic motion of the skyrmions in the PTO layer is depicted in Fig. 2(a)-2(d). It can be seen that when the tip advances 12 nm after 3 moves, the new center of the indenter has switched to a single domain, while the skyrmions on the left side move continuously to follow the tip motion (Fig. 2b). After 6 and 9 moves, the skyrmions move further to occupy the "empty" space left when the indenter moves. The front edge of the skyrmions advances about half the diameter of a skyrmion away from the initial configuration, as compared to the blue dotted line after 3 moves. After 6 moves (Fig. 2c), it can be seen that the 6 skyrmions have completely passed the blue line, indicating that they have moved at least 10 nm. After all the 9 moves (Fig. 2d), the applied stress has shifted forward by 36 nm, and the front edge of the skyrmions has moved about 24 nm. The line plots (Fig. 2e) of the local $P_x$, $P_y$, $P_z$ along the *x* direction show that the in-plane polarization is reduced to around zero and the out-of-plane polarization near the indenter is around -0.23 C·m$^{-2}$, confirming the formation of $c^-$ domain underneath the indenter after the application of the mechanical force. The average velocity of the skyrmions close to the indenter (number 1-8 in Fig. 2a) ranges from 4 m/s to 7 m/s, as plotted in Fig. 2(f). This speed is similar to the previous reports for the domain wall velocity [32], with a magnitude of ~5 m/s. This study demonstrates that: 1) the skyrmion bubbles are quasi-particles, which show topological protection where they can move collectively without breaking the structure; 2) the motion of the polar skyrmions can be activated with integrated thermal/electrical/mechanical forces.

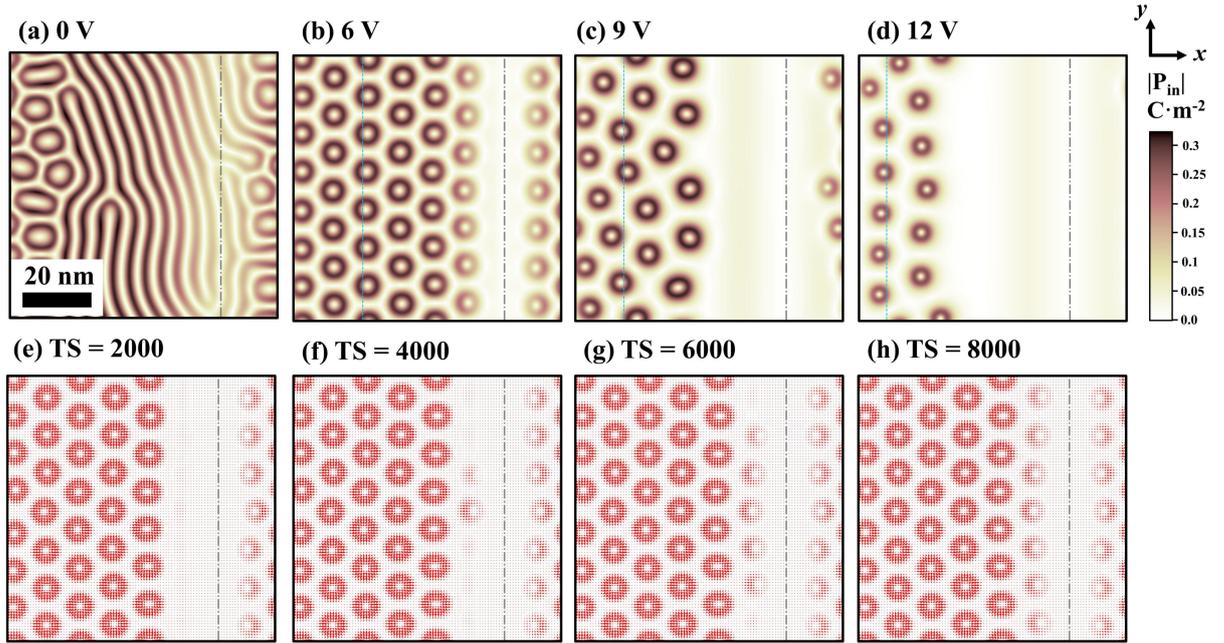

**Fig. 3| The effect of applied voltage on the skyrmion motion.** (a-d) The inplane polar structures after 9 moves under externally applied voltage of 0, 6, 9, and 12 V, respectively. (e-f) The nucleation and growth process of skyrmions under applied voltage of 6 V.

The effect of the applied bias on the motion of the skyrmions is shown in **Fig. 3**. The temperature and applied force are fixed at 500 K and 4 μN, respectively, while the applied voltage changes from 0 to 12 V. The systems exhibit vastly different polarization evolution kinetics under different bias after 9 moves, as shown in Fig. 3(a)-3(d). Without the external electric field (Fig. 3a), the 4 μN applied force could not trigger the complete switching of skyrmions, the skyrmions will merge to form static long stripes instead. Increasing the voltage to 6 V (Fig. 3b), a *c*- domain band occurs and moves with the mechanical force rightward. However, in this case, no skyrmion motion is observed, new skyrmion arrays nucleate when the tip moves. Only when the applied voltages increase to 9 V (Fig. 3c) and 12 V (Fig. 3d), the skyrmions move along the indenter moving direction, with wider *c*-domain regions underneath the indenter. The details of the nucleation and growth process of the skyrmions with a moving indenter under applied voltage of 6 V are shown in Fig. 3(e-h). Initially, after 2000 timesteps (Fig. 3e), the area underneath the tip has been fully erased to a single *c* domain with no obvious nucleation centers. Then, after 4000 timesteps, two skyrmions and several small nucleation sites can be seen in Fig. 3(f). More skyrmions nucleate and grow after 600 timesteps, as shown in Fig. 3(g). Eventually, after 8000 timesteps (Fig. 3h), a total of 7 new skyrmions nucleate and grow to fill the *c*-domain regions. This study demonstrates that the large applied voltage is required to fully decouple the skyrmions and avoid direct nucleation of skyrmions.

To further understand the impact of the applied electric field, we plot the polar structures without applied forces under different voltages at 500 K in **Fig. S2**. As the voltage increases, the in-plane polarization decreases and the shape of the skyrmions tends to be circular rather than oval in planar view. Then the line plots of the in-plane polarization (Fig. S2e), out-of-plane polarization (Fig. S2f), and the absolute value of the polarization (Fig. S2g) through the center of one skyrmion (marked by dotted box in Fig. S2a-2d) along the x direction is plotted. The two peaks in Fig. S2(e) represent the boundary of the skyrmion. As the voltage increases, the in-plane polarization value of the skyrmions decreases, and the peaks get closer gradually, showing the continuous shrinking of the polar skyrmion under high voltages. This could reduce the overall correlations between the polar skyrmions. Moreover, the moving barriers for smaller skyrmions also decrease.

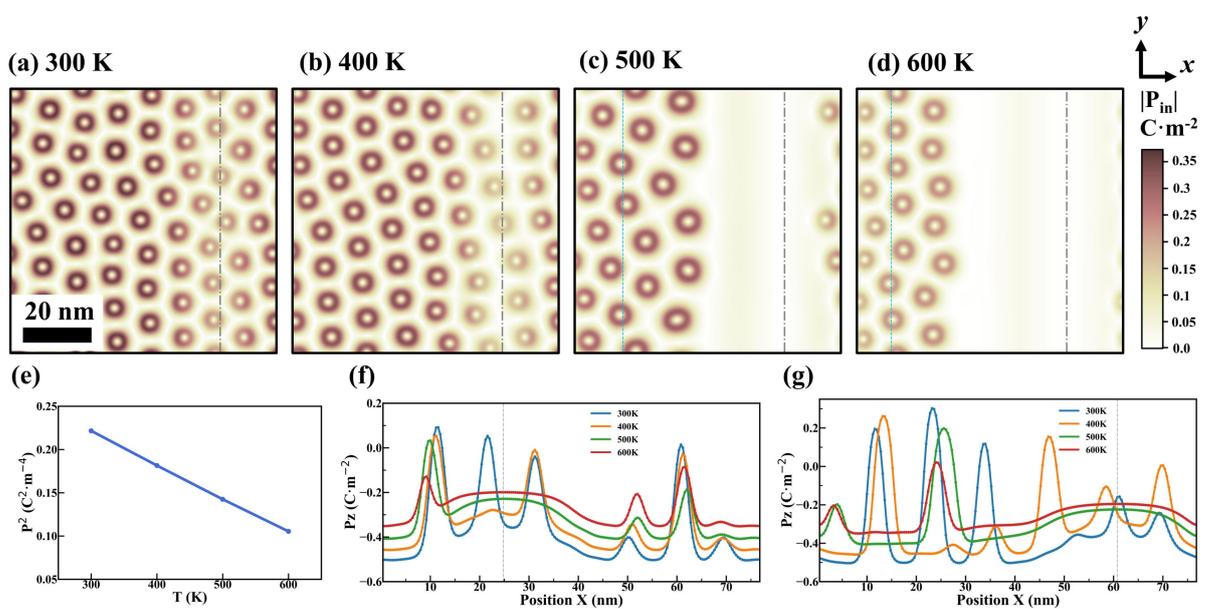

**Fig. 4| The effect of temperature on the skyrmion motion.** (a-d) The polar structures

after 9 moves under 300, 400, 500, and 600 K, respectively. (e) The average polarization square as a function of temperature without mechanical forces. (f-g) The line plots of out-of-plane polarization at different temperatures after 0 and 9 moves, respectively.

The influences of temperature on the motion of the skyrmions (applied voltage: 9 V, mechanical force: 4 μN) are explored (Fig. 4). At low temperatures (e.g., below 500 K) where the skyrmions underneath the indenter can't be fully switched to c-domain, the strong correlations between the skyrmions could hinder the skyrmion motion (Fig. 4a-b). Increasing the temperature to 500 K (Fig. 4c) and 600 K (Fig. 4d), the c-domain stripes formed and the skyrmions move rightward along with the mechanical force. As the temperature increases, the average polarization squared decreases (Fig. 4e), indicating that higher temperatures can reduce the spontaneous polarization as expected. The line plots of the out-of-plane after 0 and 9 moves are plotted in Fig. 4(f-g) respectively. At 300 K and 400 K, after applying mechanical forces, the polarization reduces while the skyrmions state is maintained. The out-of-plane polarization underneath the indenter becomes uniform, indicating the formation of a single c-domain when the temperature is above 500 K. The inplane polar structures at different temperatures without external mechanical force are shown in Fig. S3(a-d). While the line plots of the in-plane polarization, out-of-plane polarization, and the absolute value of the polarization through the center of a specific skyrmion in the x direction are given in Fig. S3 (e-g), respectively. Different from the impact of voltage, the in-plane polarization decreases with increasing temperature, while the size of the skyrmion is almost constant. Heating decreases both the in-plane and out-of-plane polarization but does not change the shape of the skyrmions. The reduced polarization under elevated temperature could lower the skyrmion motion barrier.

In addition, we investigate the impact of the moving velocity of the indenter (Fig. S4). First, we apply the mechanical forces at the same position and wait for the system to reach equilibrium. The voltage is 9V, the mechanical force is 5 μN, and the temperature is 500 K.

Then the speeds of the indenter are set to be 10000, 1000, 100, and 10 m/s in different cases, and the results after 9 moves are plotted in Fig. S4(a-d) respectively. When the indenter moves at a fast speed, i.e., >100 m/s (Fig. S4a-S4c), the tip moves too fast where the skyrmion bubble can't follow. When the moving speed of the indenter is reduced to 10 m/s, close to the skyrmion moving speed (Fig. S4d), the skyrmions move much longer than in the cases of a faster force movement velocity. This illustrates that the velocity of skyrmions is an intrinsic parameter that is dominated by the domain wall velocity.

**Conclusions**

In conclusion, we report the dynamic motion of polar skyrmions in a PTO/STO superlattice with combined mechanical, electrical, and thermal stimuli. The average moving velocity of the skyrmions is close to 5 m/s, similar to the domain wall velocity. The reduction of polarization at elevated temperatures could reduce the spontaneous polarization and hence the skyrmion motion barrier. While the applied electric field could reduce the skyrmion size and decrease strong skyrmion-skyrmion interactions. Whereas a moving mechanical indenter could switch the skyrmion to a single c-domain region that provides enough space for the skyrmions to move along. Thus, we have demonstrated that the skyrmions can move without breaking the structure, confirming the quasi-particle nature of the skyrmion bubbles, and paving the way for the practical design of skyrmion-based nanoelectronic devices such as transistors and race-track memory.

**Methods**

Phase-field simulations

The spontaneous polarization vector $\boldsymbol{P} = (P_x, P_y, P_z)$ is selected as the order parameter. The evolution of polarization is obtained by solving the time-dependent Ginzburg-Laudau equation [33]:

$$\frac{\partial P_i(\boldsymbol{r},t)}{\partial t} = -L \frac{\delta F}{\delta P_i(\boldsymbol{r},t)} \tag{Eq.1}$$

where *r* is the spatial vector, *t* is evolutionary time and *L* is the kinetic coefficient. *F* is the free energy functional of the system, consisting of four contributions, the Laudau, mechanical, electric, and gradient energy densities, integrated over the entire volume (*V*) of the film:

$$F = \int (f_{Laudau} + f_{elastic} + f_{electric} + f_{gradient})dV \quad (Eq.2)$$

Detailed expressions of the free energies and the simulation parameters can be found in previous literature [6, 8, 34, 35].

A three-dimension mesh of 192×192×350 discrete grid points is used in the simulation system, with each grid point representing 0.4 nm. The periodic boundary condition is applied in the in-plane dimensions and the superposition method is applied in the out-of-plane direction. The thickness of $(PTO)_{16}/(STO)_{16}$ superlattice film is assumed to be 120 nm with a 12 nm thickness of STO substrate, and 8 nm of air. The normalized timestep is set to be 0.01.

**Acknowledgement**

This work is supported by the National Natural Science Foundation of China (grant No. 92166104) and the Joint Funds of the National Natural Science Foundation of China (grant no. U21A2067, YW). ZH gratefully acknowledges a start-up grant from the Zhejiang University. The simulation results in this work were obtained using the Mu-PRO software package (https://muprosoftware.com). The phase-field simulation was performed on the MoFang III cluster on Shanghai Supercomputing Center (SSC).

**Declare of Interest**

The authors declare no conflict of interest.

**Data Availability Statement**

The data that support the findings of this study are available from the corresponding author upon reasonable request.

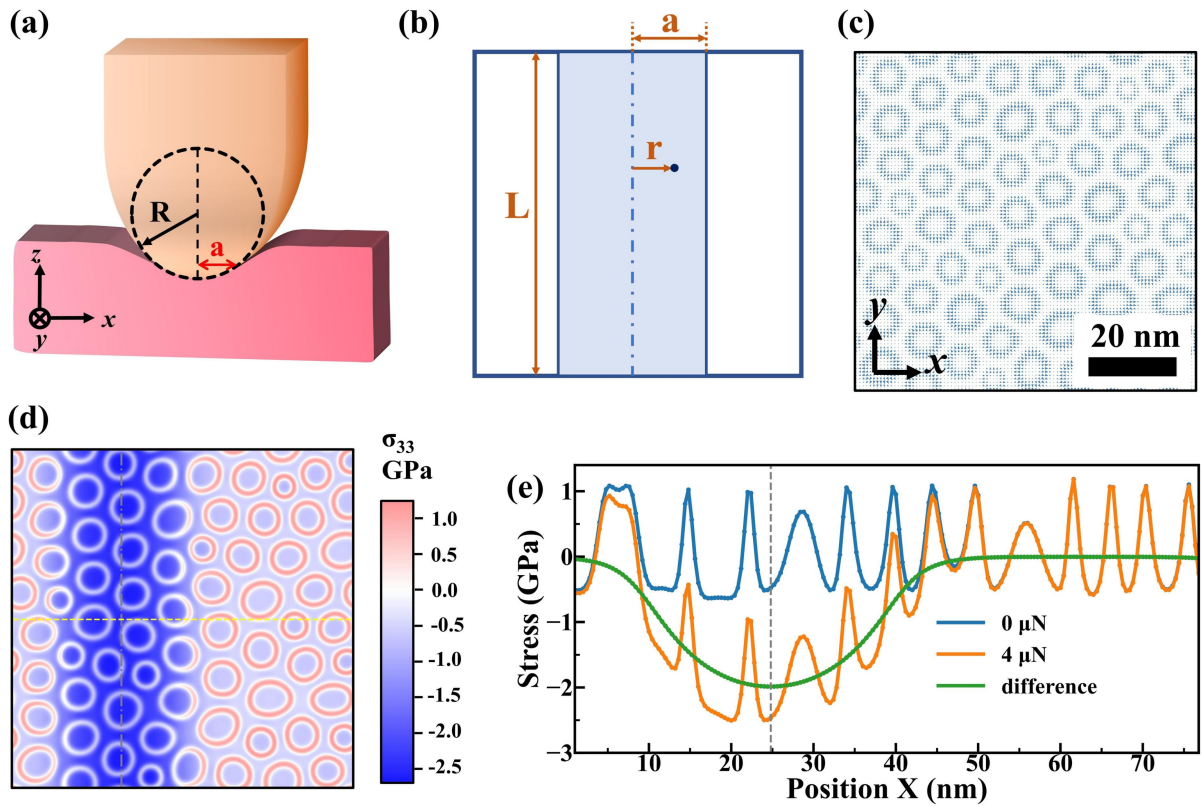

**Fig. S1| Schematics of the initial setup at 300 K and 0 V.** (a) The schematics of the indenter. (b) The 2-D surface contact between the tip and film. (c) The planar view of the skyrmions at top PTO layer without mechanical force at 300 K. (d) The stress component $\sigma_{33}$ after applying a 4 μN force at 300 K. The grey dash dot line represents the center of the banded stress. (e) The line plots of stress component $\sigma_{33}$ with and without external mechanical force along the yellow line in (d).

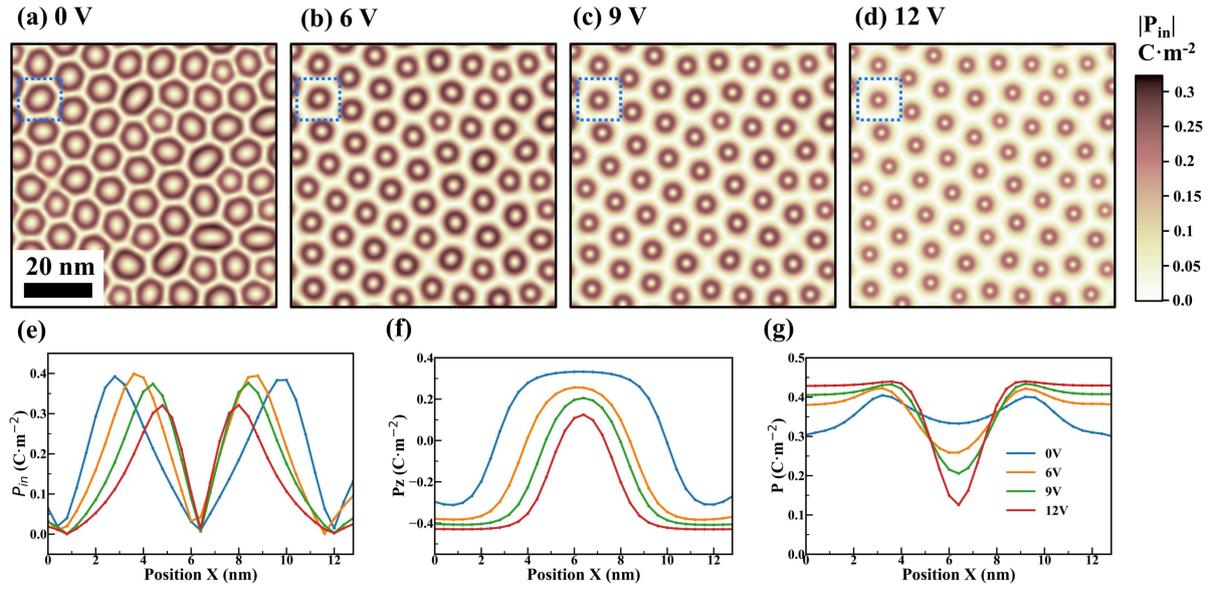

**Fig. S2| Influence of the applied voltage.** (a-d) The patterns of different polor structure after applying 0, 6, 9, and 12 V external voltage, at 500 K respectively. (e-g) The line plots of the in-plane polarization (e), out-of-plane polarization (f) and the absolute value of the polarization (g) through the center of the skyrmions in the blue dotted box in (a-d).

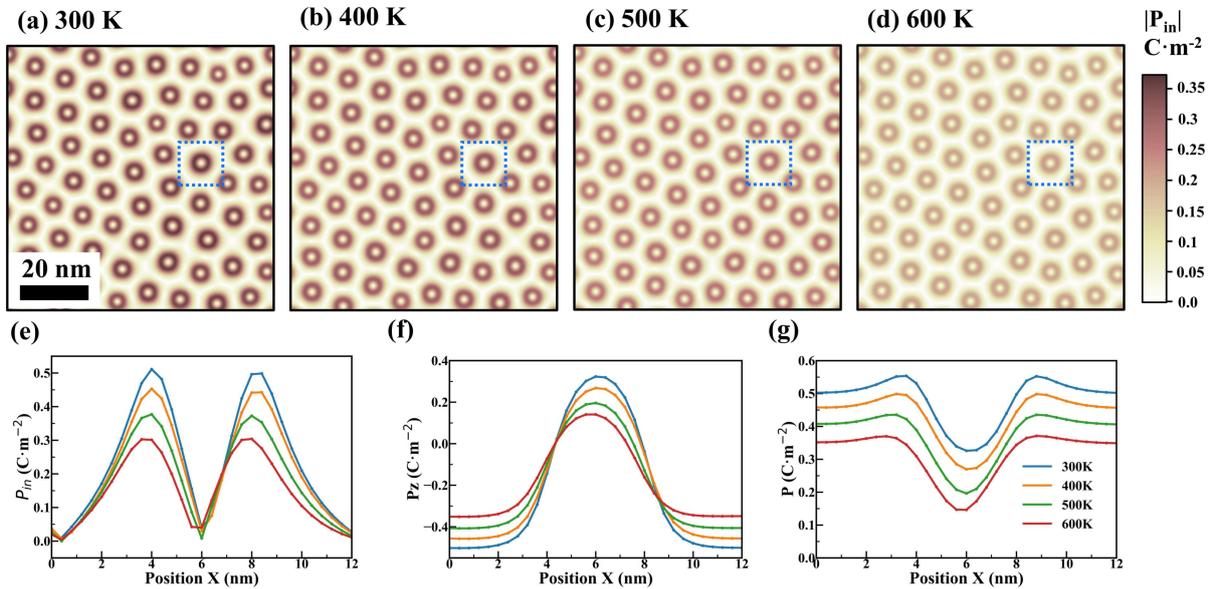

**Fig. S3| Influence of temperature.** (a-d) The patterns of different polor structure under 9 V at 300 K, 400 K, 500 K, 600 K, respectively. (e-g) The line plots of the in-plane polarization (e), out-of-plane polarization (f) and the absolute value of the polarization (g) through the center of the skyrmions in the blue dotted box in (a-d).

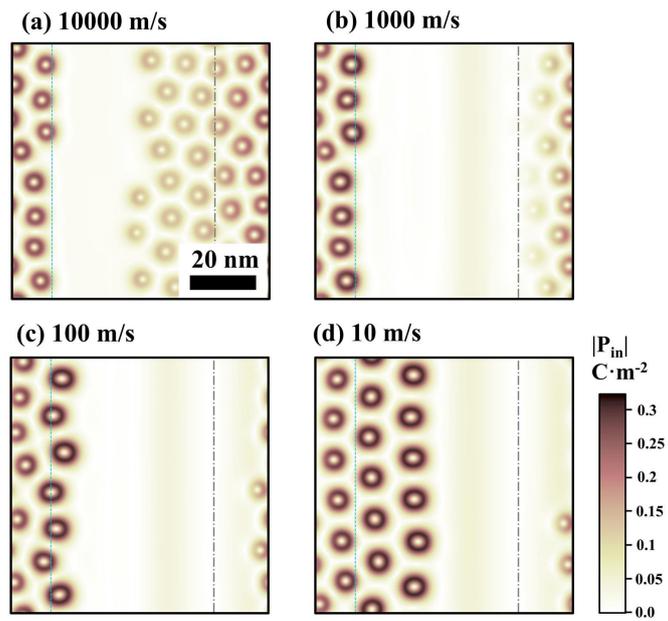

**Fig. S4| Influence of the indenter moving speed on the skyrmion moving speed.** (a-d) The patterns of different polor structure after 9 moves with the speed of the mechanical indenter be 10000, 1000, 100, and 10 m/s, respectively, under 9 V, at 500K.